# A Novel Quaternary Decoder Design Utilizing 32nm CMOS and GNRFET Technology for Enhanced High-Density Memory Applications

Anindita Chattopadhyay
*Dept. of Electronics and Communication*
*BMS College of Engineering*
Bengaluru, India
anindita.lvs21@bmsce.ac.in

Pooja Desai
*Dept. of Electronics and Communication*
*BMS College of Engineering*
Bengaluru, India
pooja.lvs21@bmsce.ac.in

Vishwas P
*Dept. of Electronics and Communication*
*BMS College of Engineering*
Bengaluru, India
vishwasp.lv21@bmsce.ac.in

Dr. Vasundhara Patel K.S
*Dept. of Electronics and Communication*
*BMS College of Engineering*
Bengaluru, India
vasu.ece@bmsce.ac.in

*Abstract*— **Multi-Valued Logic (MVL) has more than one logic level defined to represent data whereas binary logic has 2 logic levels. It has been shown that the MVL circuits use the circuit resources more effectively at different voltage levels with less circuitry and greater efficiency. Recently, graphene nano-ribbon field effect transistor (GNRFET) has drawn a lot of interest due to its higher electron mobility. This paper presents quaternary decoder implemented in GNRFET and analyzed latency, power, performance etc. also compared the power and delay characteristics of the design implemented both in CMOS and Graphene Nano Ribbon Field Effect Transistor (GNRFET) in the 32nm technology node.**

## I. INTRODUCTION

The Binary logic is widely used in the semiconductor industry. MVL circuits need optimization compared to binary number representation and ongoing research in this area is advancing in this area. MVL reduces the amount of metal connections, which lowers chip size and power consumption. The Multi-Valued Logic (MVL) includes several special features for resolving the connection problems. Three levels were first introduced by Lukasiewicz [1] ninety years ago, and a large number of logic levels have been built on his foundation. MVL is the study of discrete N-valued systems when L > 2, L is the domain of digital representation . In a broad sense, discrete variables with an infinite number of values and binary valued variables can both be regarded as MVL systems[2].

When it comes to testing of the integrated circuits (ICs), the volume of test data in ICs is rising along with their complexity[14]. The use of Multi-Valued Logic (MVL) is one approach to address test economics. Because MVL can have more than two values, MVL-based ATE can broadcast a number of test data bits simultaneously in a line. Therefore, the test data can be increased proportionally to the MVL bit steps even if the test clock speed is the same as that of the DUT.

MVL circuits can be designed in two modes: current mode and voltage mode. Current-mode CMOS Multi-Valued Logic circuits have applications in digital signal processing and computing. In this paper, a low-complexity voltage-mode multi-valued logic decoder design that translates a quaternary logic value into binary that is particularly aimed at high-density memory applications.

Two particular technologies have been chosen to design and implement the MVL decoder: CMOS and GNRFET. CMOS, because it has high noise immunity and consumes low power. GNRFET, because of its high electrical properties and low power applications. The short-channel effects that are common in CMOS smaller than 100 nm can be avoided using GNRFET. Compared to a CMOS, less Energy Delay Product (EDP) and Power Delay Product (PDP) are observed by GNRFET implementation.

Graphene-Nano Ribbon-Field-Effect-Transistor (GNRFET) is the most studied carbon-based field-effect transistors. Due to planar structure GNRFET is the most suitable alternative to silicon without the need to replace existing CMOS manufacturing technology [11] [6].

The valence band of carbon forms a group 14 elements. Carbon can be of many non-metal forms. when carbon atoms are arranged in crystalline configurations and combined it can produce a variety of hexagonal rings resembling benzene [12]. A two-dimensional layer of carbon is called graphene atoms in their steady state with a zero bandgap. A thin strip is known as graphene with a width of less than 50 nm.

The GNR structures can be made in two forms zigzag and armchair. The armchair edged GNR is metallic and it depends on the width of the ribbon, either semiconducting or metallic [11][5].

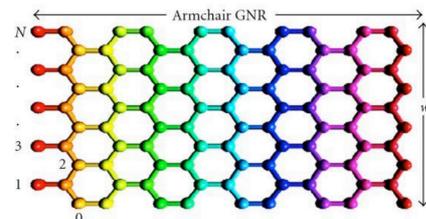

*Figure 1 – Armchair GNR width in relation to dimer lines*

High-grade graphene has excellent electron mobility and electrical conducting properties, and is strong, light, practically transparent. Its interactions with other materials and with light, as well as its inherent two-dimensional character, give unique characteristics such the bipolar transistor effect, ballistic charge transfer, and enormous quantum oscillations.

## I. QUATERNARY TO BINARY TRANSLATOR

### A. Voltage Level Converter

A circuit made up of PMOS and NMOS transistors is known as a Voltage Level Converter (VLC). To produce the necessary outputs, several threshold voltage (Vth) combinations are used. Voltage Level Converters come in three different kinds: VLC1, VLC2, and VLC3.

$$D_i(x) = \begin{cases} r-1, & \text{if } x \leq i \\ 0, & x \geq i+1 \end{cases}$$

Where, i varies from 0 to 2 for VLC1 to VLC3 respectively, $i \in \{0, 1, \ldots r-2\}$

And x is the input which is varying from 0 to 3, $x \in \{0, 1, \ldots r-1\}$

In table 1, truth tables for VLC1, VLC2, and VLC3 are displayed.

| Input(volts) | Output(volts) | | |
|---|---|---|---|
| | VLC1 | VLC2 | VLC3 |
| 0 | 3 | 3 | 3 |
| 1 | 0 | 3 | 3 |
| 2 | 0 | 0 | 3 |
| 3 | 0 | 0 | 0 |

Table 1- Truth table of VLC1, VLC2, and VLC3

### A.1 Voltage Level Converter 1 (VLC1)

The PMOS and NMOS transistors in the VLC1 circuit have a threshold voltage (Vth) of -2.2V and 0.2 V, respectively. Since NMOS will be switched off for input '0', PMOS will conduct, and 3V is sent to the output in this case. The output will be '0' for the remaining inputs, which are 1V, 2V, and 3V, where PMOS will be inactive and NMOS will be active.

The first column of table 1 will be verified by this circuit. The symbolic representation and circuit diagram are shown in Fig. 2 and Fig. 3 respectively.

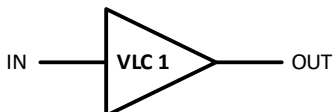

Figure 2 – Symbolic representation of VLC1

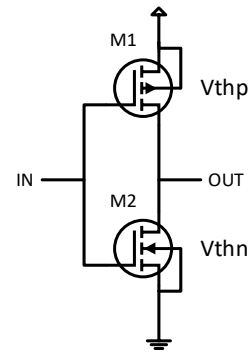

Figure 3 – Circuit diagram of VLC1

### A.2 Volatge Level Converter 2 (VLC2)

The Vth values for NMOS and PMOS transistors for VLC2 are 1.2V and -1.2V, respectively. Here, PMOS will be switched ON and NMOS will be OFF for inputs 0 and 1, resulting in an output of 3V. Transistor NMOS will conduct and transistor PMOS will be OFF for inputs 2V and 3V, hence the output will be 0.

The second column of table 1 will be verified by this circuit. The symbolic representation and circuit diagram are shown in Fig. 4 and Fig. 5 respectively.

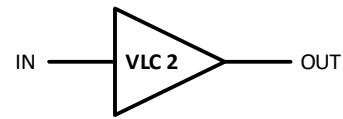

Figure 4 - Symbolic representation of VLC2

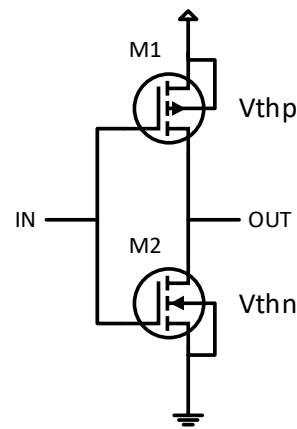

Figure 5 - Circuit diagram of VLC2

### A.3 Volatge Level Converter (VLC3)

With a Vth value of 2.2V for NMOS and -0.2V for PMOS transistors, the VLC3 circuit is built. The output will be "3V" for inputs "0," "1," and "2" because transistor PMOS will conduct and NMOS will be OFF. The output will be "0" for input "3" since the transistor's NMOS will conduct while the PMOS is OFF.

The third column of table 1 will be verified by this circuit. The symbolic representation and circuit diagram are shown in Fig. 6, and Fig. 7 respectively.

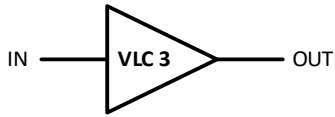

*Figure 6 - Symbolic representation of VLC3*

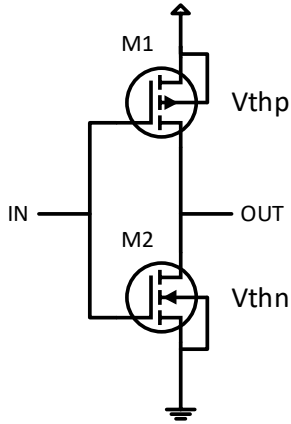

*Figure 7 - Circuit diagram of VLC3*

### B. Quternary to Binary Converter

A basic Quaternary to binary converter uses three Voltage Level converters VLC1, VLC2, VLC3, two XOR gates, and three inverters.

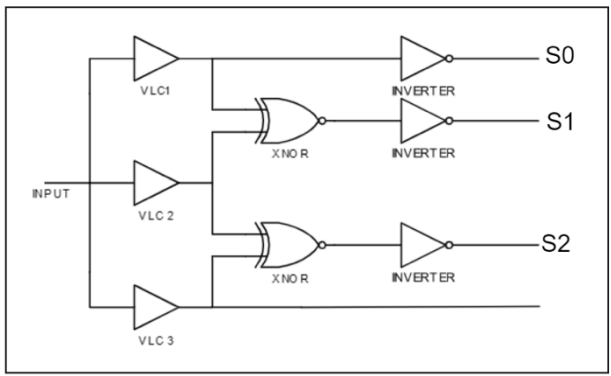

*Figure 8 - Circuit diagram of quaternary to binary converter*

The decoder circuit presented in this paper is designed and simulated on SPICE for a 32nm GNRFET and CMOS process, and the output has been observed and analyzed to compare the parameters such as power and speed. Input has been varied from 0V to 1.2V in the step of 0.4V. This circuit operates with four voltage levels corresponding to 0V and other three power supply lines of 0.4V, 0.8V and 1.2V where Logic level '0' = 0V, level '1' = 0.4V, level '2' = 0.8V, logic '3' = 1.2V.

## II. SIMULATION AND RESULT

After examining the simulation results of the circuit schematics and transient responses of the proposed work, we will now evaluate the performance of the circuits in terms of power dissipation, delay, rise time, and fall time.

The following graphs were obtained by simulating the design on H-SPICE. The graph show that the circuit could produce the desired result and the power graph has also been shown.

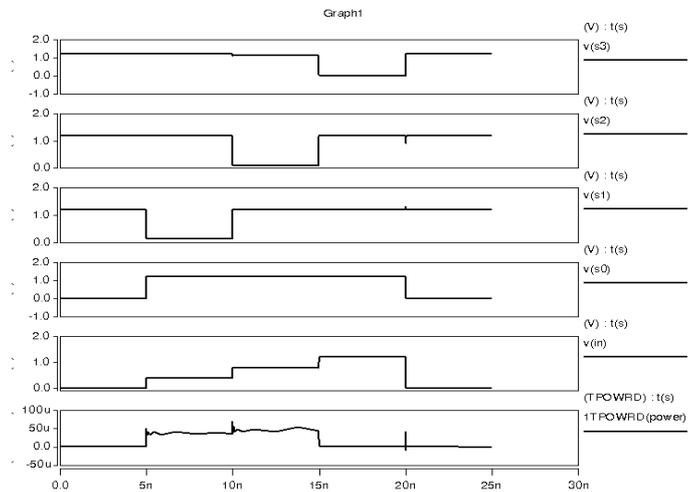

*Figure 9 – simulation result of decoder using CMOS*

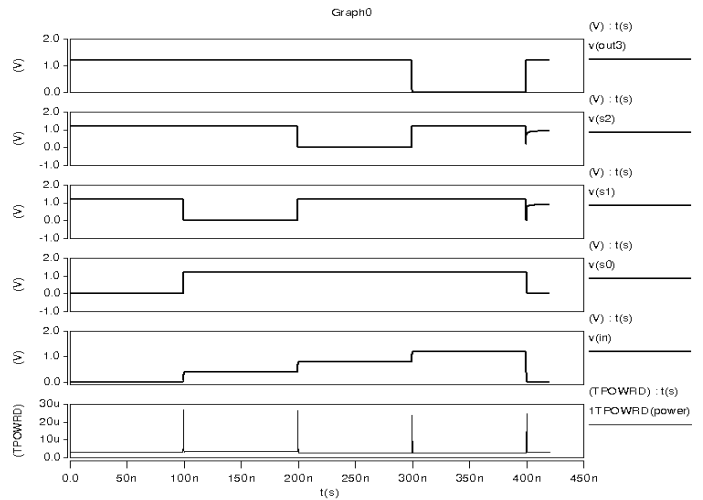

*Figure 10 – simulation result of decoder using GNRFET*

## III. ANALYSIS OF SIMULATION

It has been observed that CMOS consumes 3.0899nW power while GNRFET consumes 7.029nW. The performance delay calculation shows that GNRFET performs better than CMOS at 32nm technology. It could be said that GNRFET decoders are have improved performance than the CMOS ones.

| Technology (32nm) | Max. Power consumed (nW) | Rise time (ps) | Fall time (ps) | Performance Delay Product (PDP) |
|---|---|---|---|---|
| CMOS | 3.0899 e-06 | 174.38 | 194.4 | 100.015 |
| GNRFET | 0.7029 e-06 | 4.25 | 4.81 | 4.9973 |

## IV. CONCLUSION

In this paper, a Multi-Valued Logic decoder has been designed for high density memory applications using two different technologies: CMOS and GNRFET. Decoder designed using GNRFET shows better performance. GNRFET decoder shows 82% decrease in power, 97.56% improvement in rise time, 97.52% improvement in fall time.

As a future work one can test the circuit in the memory array.